\pdfoutput=1
\documentclass[11pt]{article}

\usepackage[final]{acl}

\usepackage{times}
\usepackage{latexsym}

\usepackage[T1]{fontenc}
\usepackage[utf8]{inputenc}

\usepackage{microtype}

\usepackage{inconsolata}

\usepackage{graphicx}

\usepackage{amssymb}        % Mathematical Symbols
\usepackage{booktabs}       % professional-quality tables
\usepackage{amsmath}

\title{VCTR: A Transformer-Based Model for Non-parallel Voice Conversion}

\author{Maharnab Saikia \\
  Independent Researcher / Assam, India \\
  \texttt{maharnabsaikia@gmail.com} \\}

\begin{document}
\maketitle
\begin{abstract}
Non-parallel voice conversion aims to convert voice from a source domain to a target domain without paired training data. Cycle-Consistent Generative Adversarial Networks (CycleGAN) and Variational Autoencoders (VAE) have been used for this task, but these models suffer from difficult training and unsatisfactory results. Later, Contrastive Voice Conversion (CVC) was introduced, utilizing a contrastive learning-based approach to address these issues. However, these methods use CNN-based generators, which can capture local semantics but lacks the ability to capture long-range dependencies necessary for global semantics. In this paper, we propose VCTR, an efficient method for non-parallel voice conversion that leverages the Hybrid Perception Block (HPB) and Dual Pruned Self-Attention (DPSA) along with a contrastive learning-based adversarial approach. The code can be found in \url{https://github.com/Maharnab-Saikia/VCTR}.
\end{abstract}

\section{Introduction}

Formally, let $X = \{x_i\}_{i=1}^{N_x}$ and $Y = \{y_j\}_{j=1}^{N_y}$ denote two unpaired datasets from the source and target speech domains, respectively. Each $ x_i \in X$ and $ y_j \in Y $ are mel-spectrogram representations of speech signals that differ in speaker identity, style, or acoustic domain. The objective of non-parallel voice conversion is to learn a mapping function $ G: X \rightarrow Y $ that transforms an input speech sample $x$ from the source domain into an output $G(x)$ that matches the characteristics of the target domain while preserving the underlying linguistic content of $x$. Since there is no one-to-one correspondence between samples in $X$ and $Y$, the learning process relies on adversarial and contrastive losses to align their feature distributions without paired supervision.

Previous voice conversion approaches use different training strategies, such as adversarial-based methods like Cycle-Consistent Generative Adversarial Networks \citep{cyclegan}, Variational Autoencoders \citep{vae}, and contrastive learning-based adversarial approaches like Contrastive Unpaired Translation \citep{cut}. These training strategies were initially introduced for unpaired image-to-image translation. Since unpaired voice conversion follows a similar domain-to-domain conversion process, they were applied to this task as well, like CycleGAN-VC \citep{cyclegan-vc}, VAE \citep{vae-vc}, and CVC \citep{cvc}. However, these methods still rely on convolutional neural network-based generators and focus primarily on training strategies. Although CNNs are effective at capturing local semantics, they lack the ability to capture long-range dependencies necessary for global semantics.

To address this issue, later works introduced the Vision Transformer into the generator, but it faced challenges related to generation difficulty and computational limitations.

Recently, ITTR \citep{ittr} was proposed to enhance the Vision Transformer, making it more effective and efficient. Inspired by its success, we propose VCTR (Voice Conversion Transformer), an voice conversion method that leverages the Hybrid Perception Block (HPB) for token mixing across different receptive fields to better utilize global semantics and Dual Pruned Self-Attention (DPSA) to significantly reduce computational complexity. Additionally, we incorporate contrastive learning for efficient one-way adversarial training.

\section{Related Works}

\subsection{Contrastive Learning for Unpaired Translation}
CUT (Contrastive Unpaired Translation) represents a significant advancement in the domain of
unpaired translation. Building on the foundational work of CycleGAN, CUT
introduced the use of contrastive learning to achieve high-quality translation between domains with
a more efficient model architecture. By reducing the need for multiple generators and discriminators,
CUT simplifies the learning process while still delivering impressive results in terms of output quality.

Later, the CycleGAN approach led to the introduction of CycleGAN-VC \citep{cyclegan-vc}. Inspired by the contrastive learning approach in CUT, Contrastive Voice Conversion (CVC) \citep{cvc} was introduced for voice conversion, achieving promising results in voice conversion tasks.

\subsection{Vision Transformer}
The Vision Transformer (ViT) \citep{vit} is a pioneering model that applies the transformer architecture, originally developed for natural language processing, to the domain of image analysis. Unlike traditional convolutional neural networks (CNNs), ViT uses self-attention mechanisms to process image patches as sequences, enabling it to model long-range dependencies across an entire image. This allows ViT to develop a comprehensive global understanding of visual data, making it particularly effective for tasks requiring complex feature extraction.

While ViT has demonstrated state-of-the-art performance on various benchmarks, particularly with large-scale datasets, other research has shown that combining CNNs with vision transformers can achieve even better results \citep{earlyconvolutionshelptransformers}. This hybrid approach leverages the strengths of both architectures: CNNs excel at capturing local features, while transformers provide global context, leading to improved performance across diverse tasks. However, it faced challenges related to generation difficulty and computational limitations. To address these issues, ITTR was later introduced, making the Vision Transformer more efficient. In our work, we employ this architecture.

\subsection{Generative Adversarial Networks}
Generative models have become foundational in the field of deep learning, enabling the creation of new data instances that resemble a given dataset. Among these, Generative Adversarial Networks (GANs) \citep{gan} have been particularly influential due to their ability to generate high-quality data. The adversarial framework, consisting of a generator and a discriminator, allows GANs to learn complex data distributions effectively. However, GANs often suffer from issues such as mode collapse, where the model generates limited diversity in outputs, and training instability, requiring careful tuning of hyperparameters and network architectures. We encountered instability during training, which is further discussed in the experiments section.

\section{Methods}

\begin{figure*}[!ht]
    \centering
    \includegraphics[width=1\linewidth]{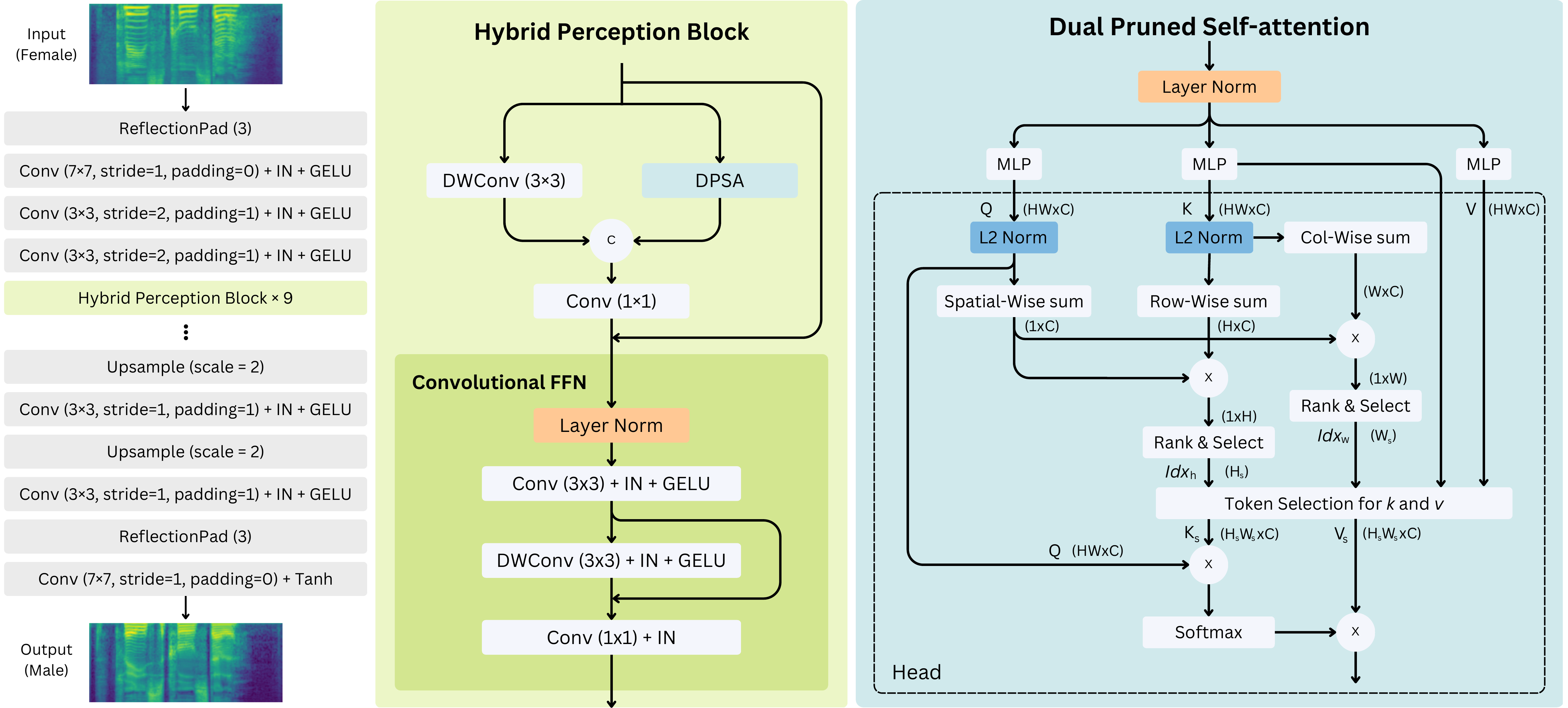}
    \caption{The VCTR architecture, consists of a ResNet on the left and a single Hybrid Perception Block (HPB) on the middle. Concatenation is denoted by 'C,' while 'IN' represents Instance Normalization and 'Conv' denotes Convolution. Dual Pruned Self-Attention (DPSA) on the right, 'L2 Norm' to indicate L2 normalization and 'X' to denote matrix multiplication. The dotted box represents operations within a single attention head. For best clarity, view the diagram zoomed in.}
    \label{fig:generator}
\end{figure*}

In this section, we introduce the architecture and detailed design of VCTR, including its overall structure, key building blocks, and the efficient approach for the Vision Transformer, namely the Hybrid Perception Block and Dual Pruned Self-Attention. Finally, we present the learning objectives.

\subsection{Model Architecture}

\subsubsection{\textbf{Generator}} The key modification in this approach is replacing the original generator in CVC with the ITTR generator architecture. The generator consists of a ResNet \citep{resnet}, similar to CycleGAN, but with a Hybrid Perception Block at the bottleneck. The Hybrid Perception Block integrates convolution for learning local semantics and Dual Pruned Self-Attention for capturing global semantics.

In our approach, we convert audio into mel spectrograms, treating them as 2D image-like representations of audio. Mel spectrograms encode both time and frequency information in a structured visual format, making them suitable for processing with image-based architectures.

As shown in Figure \ref{fig:generator}, we use the initial convolutional stem of the CycleGAN generator to produce overlapping patch embeddings of the input image. The patch embedding layer consists of three stacked convolutional layers, and the resulting overlapping patch embeddings have dimensions of 13×13 pixels.

This is followed by nine Hybrid Perception Blocks (HPBs) stacked in the bottleneck of our generator, consistent with the generator architecture in ITTR \citep{ittr} and the original CycleGAN \citep{cyclegan}. Lastly, the decoder of VCTR is implemented using three convolutional layers.

\textbf{Hybrid Perception Block:} The Hybrid Perception Block (HPB) consists of two branches designed for capturing informative features and token mixing: a local branch for extracting local features from tokens and a global branch for capturing long-range dependencies. These branches are concatenated and passed through a convolutional feedforward network.

The local branch utilizes a convolutional neural network, specifically depthwise convolution (DWConv), to reduce complexity, while the global branch employs self-attention to establish relationships between individual token pairs and their context. After concatenation, the tokens are processed through a convolutional MLP to integrate contextual information from both branches. Finally, they pass through a feedforward convolutional block for further token fusion. Instance normalization \citep{instance-norm} \citep{improvedtexturenetworksmaximizing} and GELU \citep{gelu} activation are applied to stabilize the flow in the convolutional feedforward network.

\textbf{Dual Pruned Self-Attention:} DPSA technique as shown in Figure \ref{fig:generator}, reduces computation and memory usage by focusing only on the most important tokens before calculating attention. This avoids the costly process of computing attention between every pair of tokens. Instead of analyzing each token individually, tokens are grouped into rows and columns. The contribution of each row and column to the attention is calculated, and only the top rows and columns with the highest contributions are retained. Attention is then computed only on these pruned tokens.

The token scoring is formulated as follows by grouping them into rows and columns and summing their values across each row and column:

\small \begin{align}
    Score_r &=\sum_{i = 1}^{N}\sum_{j = 1}^{W}q_ik_{rj}^T  \notag \\ 
            &= \left(\sum_{i = 1}^{N}q_i\right) \left(\sum_{j = 1}^{W}k_{rj}\right)^T, \quad r \in \{1, \dots, H\}
\end{align}

\small \begin{align}
    Score_c &= \sum_{i = 1}^{N}\sum_{j = 1}^{H}q_i k_{jc}^T  \notag \\ 
            &= \left(\sum_{i = 1}^{N}q_i\right) \left(\sum_{j = 1}^{H}k_{jc}\right)^T, \quad c \in \{1, \dots, W\}
\end{align}
\normalsize

Rows and columns are ranked based on their scores, and the top $N_s$ rows and $N_s$ columns are selected. $N_s$ is a hyperparameter set to $\sqrt{H}$, where $H$ is the height of the input grid. Tokens outside these selected rows and columns are discarded. An $ArgMaxScore$ operation is applied:

\small \begin{equation}
    Index_r = ArgMaxScore(Score_r)[:N_s]
\end{equation}

\small \begin{equation}
    Index_c = ArgMaxScore(Score_c)[:N_s]
\end{equation}

\small \begin{equation}
    K_s = K[Index_r, Index_c], V_s = V[Index_r, Index_c]
\end{equation}
\normalsize

Next, the pruned keys $K_s$ and values $V_s$ are reshaped into a smaller matrix, and attention is computed only between the queries $Q$ and the pruned $K_s$ and $V_s$. Temperature scaling is not used, as pruning naturally limits attention values, eliminating the need for the usual $\frac{1}{\sqrt{D}}$ factor.

\small \begin{equation}
    Attention = Softmax(Q \cdot K_s^T) \cdot V_s
\end{equation}
\normalsize

By applying L2 normalization to keys and queries, attention scores remain between -1 and 1, preventing skewed results. Since $N_s$ is set to $\sqrt{H}$ in practice, for a single attention head, the computational complexity is reduced from $O(N^2C)$ to $O(NHC)$.

\subsubsection{\textbf{Discriminator}} For the discriminator, the PatchGAN architecture is used, as described in the original Pix2Pix paper \citep{pix2pix}. The PatchGAN discriminator evaluates the realism of overlapping image patches rather than the entire image. This allows the discriminator to focus on high-frequency details. Since we treat spectrograms as 2D images, this helps to improve the sharpness and overall quality of the generated spectrograms.

\begin{figure}[ht]
    \centering
    \includegraphics[width=1\linewidth]{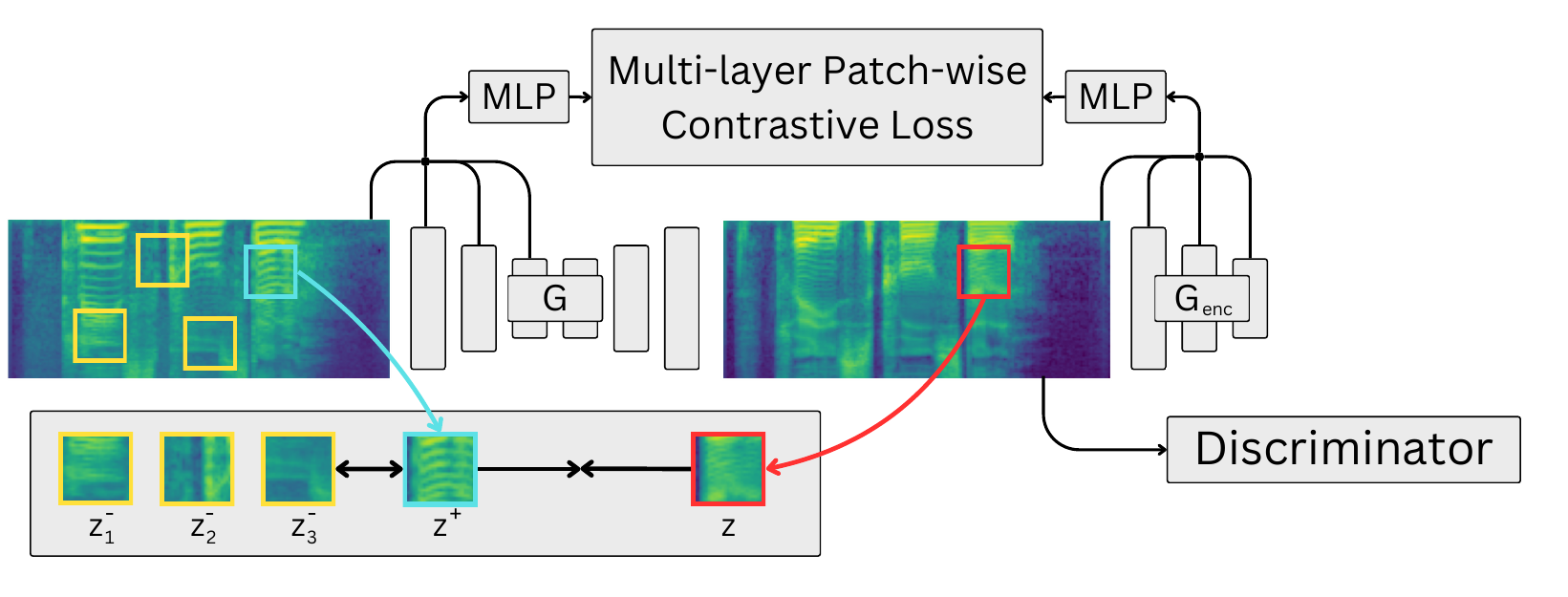}
    \caption{Patchwise contrastive learning is used for one-sided translation, A generated output patch should be closer to its corresponding input patch compared to other random patches. To achieve this, we use a multi-layer patch-wise contrastive loss, which maximizes mutual information between corresponding input and output patches.}
    \label{fig:patchnce}
\end{figure}

\subsection{Loss Functions}

\textbf{Adversarial Loss:} The adversarial loss \citep{gan}, denoted as $L_{GAN}$, encourages the generator to produce images that are indistinguishable from those in the target domain. The loss is formulated as follows:

\small \begin{align}
    L_{GAN}(G, D, X, Y) &= \mathbb{E}_{y \sim Y} \log D(y)  \notag \\ 
                         &\quad + \mathbb{E}_{x \sim X} \log (1 - D(G(x)))
\end{align}
\normalsize

where $G$ is the generator, $D$ is the discriminator and $X$ and $Y$ are the input and target domains, respectively.

\textbf{Patchwise Contrastive Loss:} A key component of the CUT framework. It encourages the corresponding patches between the input and output images to share similar features, thereby enhancing the consistency of the image translation. For each selected layer $l$ in the generator's encoder. The PatchNCE loss is given by:

\small \begin{equation}
l(v, v^+, v^-) = -\log{ \Bigg( \frac{\exp (\frac{v \cdot v^+}{\tau})}{\exp (\frac{v \cdot v^+}{\tau}) + \sum_{n=1}^N \exp (\frac{v \cdot v_n^-}{\tau})} \Bigg) }
\end{equation}

\small \begin{equation}
    L_{PatchNCE}(G, H, X) = \mathbb{E}_{x \sim X} \sum_{l = 1}^{L} \sum_{s = 1}^{S_l} l(\hat{z}_l^s, z_l^s, z_l^{S \setminus s})
\end{equation}
\normalsize

In this equation, $\hat{z}_l^s$ represents the features of the output patches, $z_l^s$ represents the corresponding input patches, and $z_l^{S \setminus s}$ represents negative patches from the same input image. Corresponding patches from the input and output features are pulled closer, while non-corresponding patches from the same input image are pushed apart, as illustrated in Figure \ref{fig:patchnce}.

\textbf{Final Objective:} The final loss is a combination of the adversarial loss and the PatchNCE loss $L_{PatchNCE}(G, H, X)$. There is also a PatchNCE loss $L_{PatchNCE}(G, H, Y)$, which acts as an identity loss, as explained in the CUT paper \citep{cut}. The overall loss function is expressed as:

\begin{align}
    L_{total} &= L_{GAN}(G, D, X, Y)  \notag \\ 
              &\quad + \lambda_X L_{PatchNCE}(G, H, X)  \\ 
              &\quad + \lambda_Y L_{PatchNCE}(G, H, Y) \notag
\end{align}

The weights $\lambda_X$ and $\lambda_Y$ are hyperparameters that control the contribution of each loss term in PatchNCE loss $L_{PatchNCE}(G, H, X)$ and $L_{PatchNCE}(G, H, Y)$. When using identity loss, the values are chosen as $\lambda_X = 1$ and $\lambda_Y = 1$. When identity loss is not used, the values are set as $\lambda_X = 10$ and $\lambda_Y = 0$.

\section{Experiments}

\subsection{\textbf{Non-Parallel Voice Conversion}}

\begin{table*}[ht] \small
    \centering
    \begin{tabular}{l|c|c|c}
         \toprule
         Method     & One to One                    & Many to One          & Many (unseen) to One \\
                    & (M-F) \quad (F-M)             & (M-F) \quad (F-M)    & (M-F) \quad (F-M) \\
    
         \midrule
         VAE        & 0.805 \quad 0.874             & 0.803 \quad 0.845    &   -   \quad   -   \\
         CycleGAN   & 0.925 \quad 0.951             & 0.926 \quad 0.968    & 0.910 \quad 0.935 \\
         CVC (CUT)  & 0.929 \quad 0.952             & 0.937 \quad 0.974    & 0.925 \quad 0.945 \\
         CNEG-VC    & 0.934 \quad 0.963             & 0.945 \quad \textbf{0.976}    & 0.937 \quad \textbf{0.957} \\
         VCTR (CUT) & \textbf{0.963} \quad \textbf{0.973}    & \textbf{0.946} \quad 0.967    & \textbf{0.950} \quad 0.949 \\
         \bottomrule

    \end{tabular}
    \caption{Comparison of voice similarity scores for different methods. Higher values indicate better performance.}
    \label{tab:sim_scores}
\end{table*}

\textbf{Dataset:} Experiments are conducted on the VCTK corpus \citep{vctk}, which contains 44 hours of speech data uttered by 109 native English speakers with various accents. For one-to-one voice conversion, speech data was collected from two different speakers. Fifty speech samples from the source speaker were excluded from training for evaluation.

For many-to-one voice conversion, data from 100 speakers was used, treating them as a single source domain, while 9 speakers were excluded for evaluation on unseen source speakers. Additionally, 50 unique speech samples were set aside for evaluation in the many-to-one setting.

\textbf{Training Strategy:} To ensure a fair comparison, we adopt the same training strategy as CVC. From the dataset, we extracted speech data and used only 2-second segments. No padding was applied, as described in CVC, since it can lead to model collapse.

For the vocoder, Parallel WaveGAN \citep{parallel-wavegan} was used to synthesize waveform signals. The speech data, originally in 48 kHz, was downsampled to 24 kHz. We extracted 80-band mel spectrograms and followed the default settings of Parallel WaveGAN. The window size, FFT size, and hop size were set to 1024, 1024, and 256, respectively. The frequency range was set with fmin = 80 Hz and fmax = 7600 Hz.

We used the Adam \citep{adam} optimizer with a learning rate of 2e-4 for the first 850 epochs, which was then linearly reduced to 0 over the next 150 epochs, for a total of 1000 epochs. Training was conducted on a single GPU with a batch size of 1. Since our architecture differs, we used layers [0, 4, 7, 10, 14] from the encoder of our generator to calculate PatchNCE loss.

\textbf{Evaluation:} We used a voice encoder system to measure the similarity between the generated fake voice and the real voice in the target domain. Specifically, we used Resemblyzer, an open-source speaker verification system that derives high-level voice representations using a deep learning model.

Given a speech sample, the Resemblyzer generates an embedding vector. Then, cosine similarity is computed between the generated embedding and the target embedding, producing a numerical score between 0 and 1, where 0 indicates completely different speakers and 1 signifies the same speaker.

\textbf{Baseline:} We compared VCTR with previous non-parallel voice conversion methods, as shown in Table \ref{tab:sim_scores}. It achieved comparable or better voice similarity between the generated fake voice and the real voice. The baseline methods include VAE \citep{vae-vc}, CycleGAN-VC3 \citep{cyclegan-vc3}, CVC \citep{cvc}, and CNEG-VC \citep{cneg-vc}.

Additionally, Table \ref{tab:param} presents the MACs (Multiply-Accumulate Operations) and parameters of the generator. Although VCTR is not a lightweight architecture, it achieved better results with lower complexity than previous models.

\begin{table}[ht] \small
    \centering
    \begin{tabular}{l|c|c}
         \toprule
         Method       & MACs(G)    & Params(G) \\
    
         \midrule
         VAE          & 1.1        & 1.1 \\
         CycleGAN-VC  & 12.8       & 11.4 \\
         CVC          & 12.8       & 11.4 \\
         CNEG-VC      & 12.8       & 11.4 \\
         VCTR         & 10.2       & 8.5 \\
         \bottomrule

    \end{tabular}
    \caption{Statistics of generator MACs and parameters for each method.}
    \label{tab:param}
\end{table}

While CVC used replication padding without output normalization to mitigate mode collapse, they stated that using reflection padding or zero padding led to mode collapse.

\begin{figure}[ht]
    \centering
    \includegraphics[width=1\linewidth]{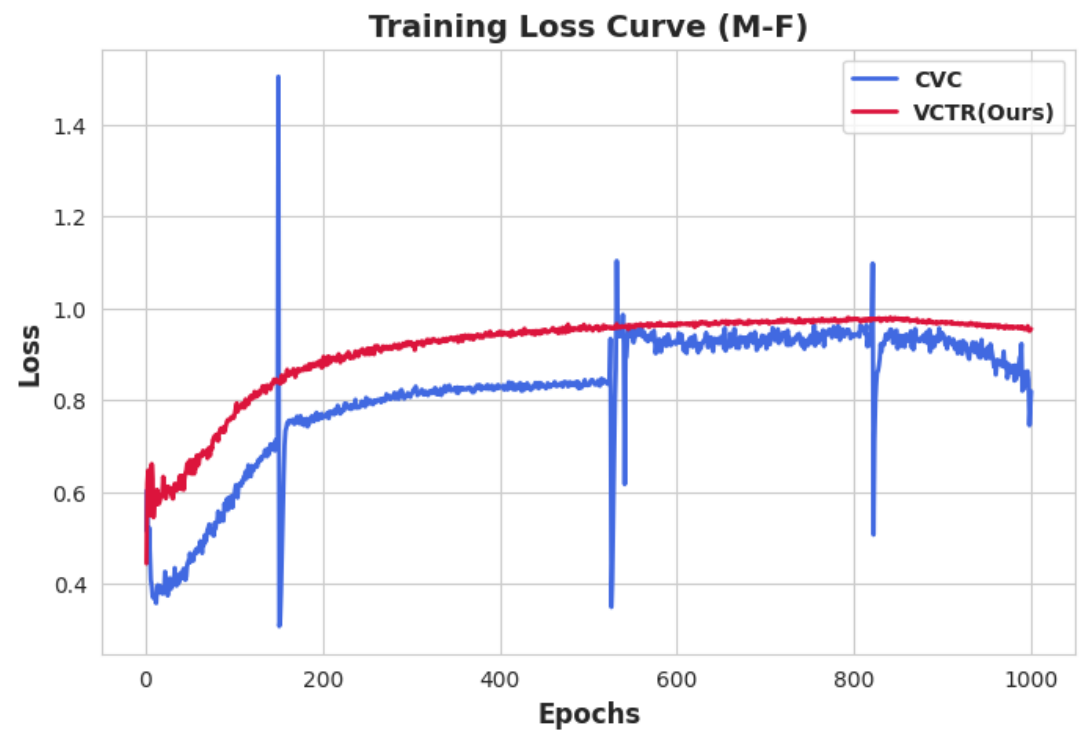}
    \caption{The loss curve from generator training is shown, with CVC represented in royal blue and our model, VCTR, in crimson. The graph shows loss spikes during male-to-female (one-to-one) voice conversion training.}
    \label{fig:loss_curve}
\end{figure}

During training, we applied different padding strategies across the entire generator. However, we still observed occasional loss spikes, indicating that the issue was not related to padding. To address this, we experimented with various techniques, such as gradient clipping. Ultimately, we found that using a Tanh activation function at the output resolved instability, even when using reflection padding, as shown in Figure \ref{fig:loss_curve}. Despite this improvement, mode collapse remains a persistent challenge in many-to-one scenarios.

We applied min-max normalization to scale the values to the range [0,1] and then adjusted the range to [-1,1] by transforming the normalized values.

\subsection{\textbf{Ablation Study}}

First, we removed each individual branch from HPB separately. Removing either the local perception branch or the DPSA branch lowered the voice similarity score, suggesting that both DPSA for global context and depth-wise convolution for local features play important roles in performance.
Next, we removed token-wise L2 normalization for Q and K, which was used to eliminate the negative impact of peaked token vectors before the softmax. This led to a significant drop in the score.
The experimental results for the ablation study are shown in Table \ref{tab:config_ablation}.

\begin{table}[ht] \small
    \centering
    \begin{tabular}{l|c|c}
         \toprule
         Configuration         & One-to-One (M-F)    & One-to-One (F-M) \\
    
         \midrule
         Ours                  & \textbf{0.959}      & \textbf{0.970} \\
         w/o DPSA              & 0.945               & 0.961 \\
         w/o Local Perception  & 0.936               & \textbf{0.970} \\
         w/o L2 Norm           & 0.937               & 0.948 \\
         \bottomrule

    \end{tabular}
    \caption{Experimental results for ablation study. Higher scores are shown in bold.}
    \label{tab:config_ablation}
\end{table}

\section{Conclusion}

In this paper, we proposed an efficient generator architecture combined with a contrastive learning method for non-parallel voice conversion. Our approach captures local features using CNNs and global context using Transformers in non-parallel voice conversion. Experimental results have shown that our method achieved better voice similarity scores compared to baseline methods.

\section*{Limitations}

However, mode collapse persists in many-to-one voice conversions. While Tanh activation stabilizes training by constraining outputs, residual issues remain. Future work could explore adaptive thresholds or spectral diversity losses to fully address these challenges. Additionally, A key limitation of the proposed architecture is its requirement for fixed-length input and output audio, which restricts the model’s ability to accommodate natural variations in speaking rate across different speakers.

\clearpage

\bibliography{custom}

\end{document}